\documentclass[reprint,amsmath,amssymb,aps,prl]{article}

\usepackage{graphicx}
\usepackage{dcolumn}
\usepackage{bm}
\usepackage{hyperref}
\usepackage[usenames,dvipsnames]{color}
\usepackage[version=3]{mhchem}
\usepackage[exponent-product = \cdot]{siunitx}
\usepackage{todonotes}
\usepackage[babel]{csquotes}

\begin{document}


\title{Modeling gel swelling equilibrium in mean-field: From explicit models to Poisson-Boltzmann}

\author{Jonas Landsgesell\thanks{Institute for Computational Physics, University of Stuttgart, D-70569 Stuttgart, Germany} \and David Sean \and Patrick Kreissl \and Kai Szuttor \and Christian Holm }

\newcommand{\TODOi}[1]{{\color{red}{#1}}}	

\newcommand{\Rend}[0]{R_\mathrm{e}}
\newcommand{\Req}[0]{R_\mathrm{eq}}

\newcommand{\DEBYE}[0]{\lambda_\mathrm{D}}
\newcommand{\kT}[0]{k_\mathrm{B}T}
\newcommand{\LB}[0]{\ell_\mathrm{B}}
\newcommand{\cs}[0]{c_\mathrm{salt}^b}
\newcommand{\pH}[0]{\text{pH}}
\newcommand{\pKa}[0]{\mathrm{p}K_\mathrm{a}}
\newcommand{\Rout}[0]{R_\mathrm{out}}
\newcommand{\Veq}[0]{V_\mathrm{eq}}
\newcommand{\Pres}[0]{P_\mathrm{res}}
\newcommand{\Pin}[0]{P_\mathrm{in}}
\newcommand{\Pcap}[0]{P_\mathrm{cap}}
\newcommand{\Pside}[0]{P_\mathrm{side}}

\newcommand{\mugel}[0]{\mu^\mathrm{gel}_{i}}
\newcommand{\mures}[0]{\mu^\mathrm{res}_{i}}
\newcommand\es[0]{\textsf{ESPResSo}\xspace}
\newcommand\PPPM[0]{{P$^3$M}\xspace}

\date{\today}
\begin{abstract}
  We develop a double mean-field theory for charged macrogels immersed in
  electrolyte solutions in the spirit of the cell model approach.  We
  first demonstrate that the equilibrium sampling of a single
  explicit coarse-grained charged polymer in a cell yields accurate
  predictions of the swelling equilibrium if the geometry is suitably
  chosen and all pressure contributions have been incorporated
  accurately. We then replace the explicit flexible chain by a suitably
  modeled penetrable charged rod that allows to compute all pressure
  terms within the Poisson-Boltzmann approximation. This model, albeit
  computationally cheap, yields excellent predictions of swelling
  equilibria under varying chain length, polymer charge fraction, and
  external reservoir salt concentrations when compared to
  coarse-grained molecular dynamics simulations of charged
  macrogels. We present an extension of the model to the
  experimentally relevant cases of pH-sensitive gels.
\end{abstract}
\maketitle

Polyelectrolyte gels consist of crosslinked charged polymers
(polyelectrolytes) that can be synthesized with various topologies and
are produced in sizes ranging from nanometers (nanogels) up to centimeters
(macrogels)~\cite{tanaka80a,harland92a}. 
They show a large, reversible uptake of water that is
exploited in numerous daily-life products, such as in superabsorbers,
cosmetics, pharmaceuticals~\cite{peppas00a,jia09a,jagurgrodzinski09a},
agriculture~\cite{zohuriaanmehr10a,kazanskii92a}, or quite recently water
desalination~\cite{hoepfner13b,richter17a}.
Tailoring polyelectrolyte gels to their applications requires a sufficiently accurate
prediction of their swelling capabilities and elastic responses, a
task that still goes beyond analytical approaches 
\cite{flory43a,katchalsky55a,khokhlov93a,rubinstein96a,claudio09a,longo11a,quesadaperez11b,jha11a,liu15a}.
So far only all-atom simulations of short single chains in the bulk (not of whole hydrogels) with explicit water have
been performed~\cite{muellerplathe97a, tonsing01a,walter10a,kosovan13a}. On the other hand, coarse-grained
polyelectrolyte network models have demonstrated their ability to amend
analytical approaches, showing that structural microscopic details can have
noticeable effects on the macroscopic properties such as the swelling
\cite{schneider02a,yan03a,
edgecombe04a,yin05a,mann05a,mann06a,yin08a,quesadaperez11a,kosovan15a,rud17a}.
Macroscopic gels with monodisperse chain length can be simulated with microscopic detail
using molecular dynamics (MD) simulations with periodic boundary conditions (PBCs) (cf.~\textit{periodic gel model}) where a unit gel section
is connected periodically to yield an infinite gel without boundaries.
However, even MD simulations of periodic gels remain computationally
very expensive due to the many particles and the slow relaxation times
of the involved polymers. Thus, the development of computationally efficient mean-field models
capable of predicting swelling equilibria have been of scientific
interest in the last years
\cite{mann05d,longo11a,kosovan15a,rud17a}. First ideas of using a
Poisson-Boltzmann (PB) cell model under tension were put forward
by Mann for salt-free gels, with moderate
success~\cite{mann05d}.

About sixty years ago Katchalsky and Michaeli~\cite{katchalsky55a}
suggested a free energy model that has recently been
shown to predict swelling equilibria reasonably well
\cite{kosovan15a} when compared to MD simulations of
charged bead-spring gels. This model has been applied to explore a
wide parameter space in search of optimal desalination
conditions~\cite{richter17a}.  However, the Katchalsky model fails~\cite{kosovan15a} for
Manning parameters
$\xi=\lambda_B/\langle d \rangle>1$~\cite{manning69a}, where $\lambda_B$ denotes the
Bjerrum length, and $\langle d \rangle$ the average
distance between polymer backbone charges. This is presumably due to the usage of the Debye-H\"uckel approximation.
%

\begin{figure}
\centering
\includegraphics[width=.7\textwidth]{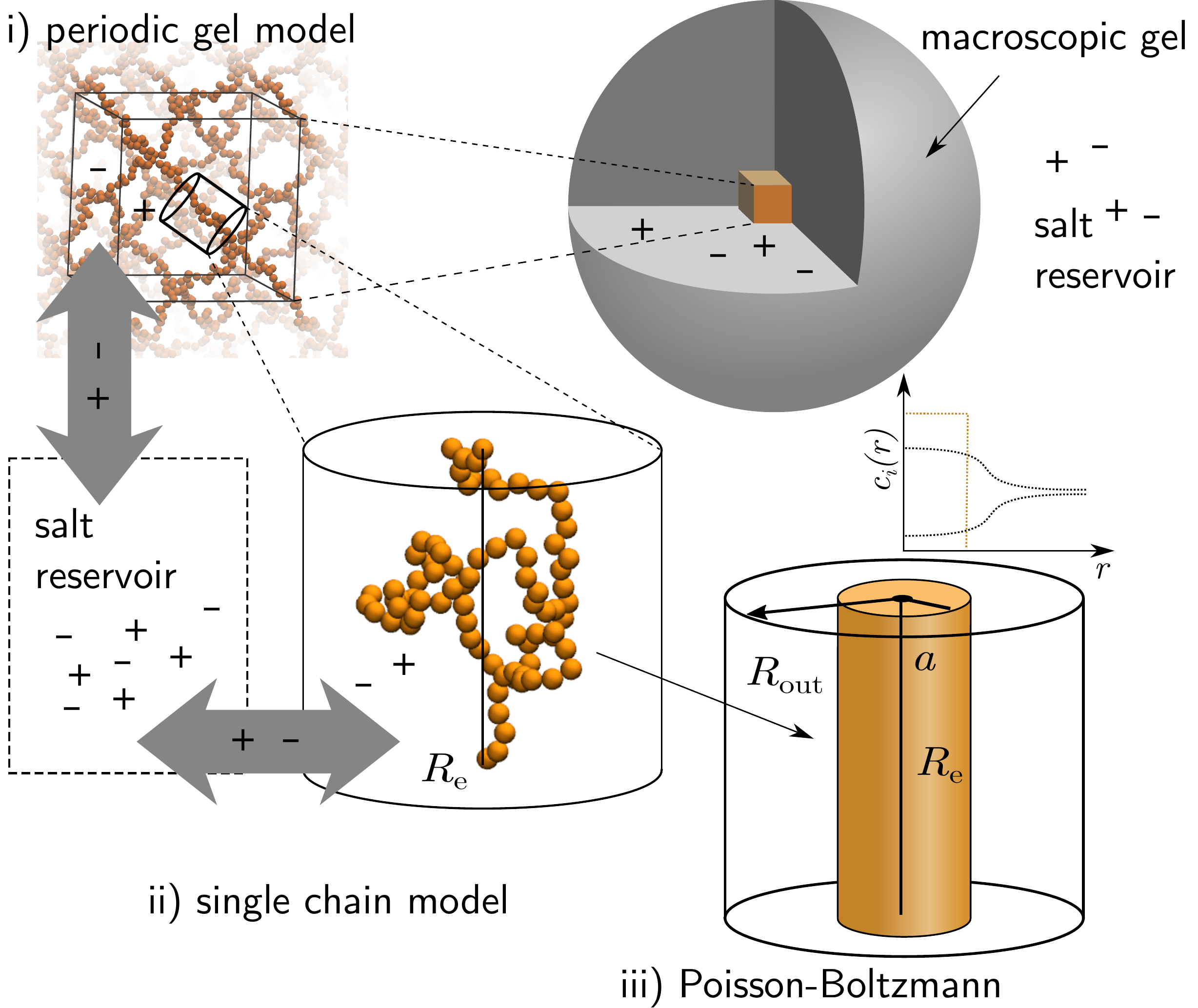}
\caption{\label{fig:schematic}(Color online) A schematic of the i) macroscopic gel; ii) single-chain; and iii) PB model of a macroscopic gel in contact with a reservoir. The plot sketches a typical radial density profile.}
\end{figure}

In this letter we describe two successive mean-field approaches to
render the determination of swelling equilibria of polyelectrolytes
accurately and efficiently. Figure \ref{fig:schematic}
displays our construction scheme of the two different models.
First, we describe a \textit{single-chain MD cell model}, that
reproduces results similar to those obtained from expensive MD
simulations of multiple crosslinked chains.
This reduces the many-body problem of the macroscopic
gel to one of computing the pressure exerted within a cell containing a single polyelectrolyte
chain under varying environmental conditions.  The
single-chain cell model can thus be viewed as a mean-field
attempt to factorize the many-body partition function of the macrogel
into a product state of suitable identical subunits~\cite{deserno01c}.
We then show that the single-chain cell model can further be
simplified in a second mean-field step using a PB description of the chain with appropriate
boundary conditions.
The PB cell description has been successful in describing a variety of
polyelectrolyte phenomena
\cite{alfrey51a,fuoss51a,katchalsky53a,deserno00a,deshkovski01a,antypov06a}
and is here applied to macroscopic polyelectrolyte gels for the first time.
The quality of our two mean-field models is gauged by comparing
them to 60 data points for the swelling equilibrium of periodic monodisperse
gel MD simulations obtained within a wide range of system parameters. 
We want to emphasize that none of our models assumes a specific stretching state of the chains: 
they are constructed to incorporate the main physical effects which happen during stretching (at high chain extensions) and compression of a polyelectrolyte gel (at low chain extensions).

Finally, we generalize the PB cell model to account for the effect of
weak groups along the chain backbone.  From this we can
efficiently predict swelling equilibria for the
experimentally relevant cases of weak polyelectrolyte gels.

We specifically compare our results to MD data for the \textit{periodic gel model}
obtained by Ko{\v s}ovan et. al~\cite{kosovan15a} where a continuum
solvent is used with standard charged bead-spring polymers
connected in a diamond lattice as
in the work of Ref.~\cite{mann05a} together with explicit salt and counterions.
Here a perfect tetrafunctional gel is described by the chain length $N$ and the monomers charge fraction $f$.
This explicit particle based model uses monodisperse
chain lengths which is in contrast to the heterogeneity observed in a wide variety of synthesized gels
\cite{panyukov96b}.  It would be computationally very costly to
introduce chain length heterogeneity into this model since it would require to simulate
a much larger representative volume element. 
However, these periodic gel simulations of a monodisperse gel are sufficient to test the validity of our two consecutive mean-field approaches.

All MD simulations (namely the periodic gel and the single-chain model described later)
are performed with PBCs using the simulation package
ESPResSo~\cite{limbach06a, weik19a}. All particles interact via WCA
interactions~\cite{weeks71a,slater09b}. Monomers
are connected via FENE bonds (including the ends of the single
periodic chain) with Kremer-Grest parameters~\cite{grest86a}.  We
employ the Langevin thermostat~\cite{frenkel02b}, and all
electrostatic interactions between particles are calculated with the
P3M method~\cite{deserno98a} tuned to an accuracy in the root mean
squared error of the electrostatic force of at least $10^{-3}$ in
electrostatic simulation units~\cite{deserno98b}.


In addition, salt ion pair exchanges between the simulation volume and an
external reservoir are performed using grand canonical Monte Carlo
moves~\cite{frenkel02b}. The equilibrium pressure
inside the gel and the electro-chemical potentials of all $i$ species
balance out with that of the reservoir: $\Pin(\Veq)=\Pres$ and
$\mugel=\mures,$ respectively. The simulations are performed at
different imposed volumes and the internal pressure is
measured after chemical equilibrium is reached. 
The reservoir pressure is approximated by the ideal gas expression $\Pres=\sum_i \kT c_i^\mathrm{b}$,
with the Boltzmann constant $k_\mathrm{B}$, temperature $T$ and
bulk ion concentrations $c_i^\mathrm{b}$, which are chosen such
that the bulk is electroneutral.

The above approach requires multiple simulations at different imposed volumes until
the condition $\Pin(\Veq)=\Pres$ can be narrowed down to a
satisfactory small interval.
The equilbrium volume is found at the intersection of $\Pin(\Veq)$ and $\Pres$ by using linear interpolation. The errorbar is given by the width of the interval. Under
equilibrium conditions, the end-to-end distance $\Rend$ is equal to
the equilibrium chain extension $\Req$.

\begin{figure}
\includegraphics[width=.8\columnwidth]{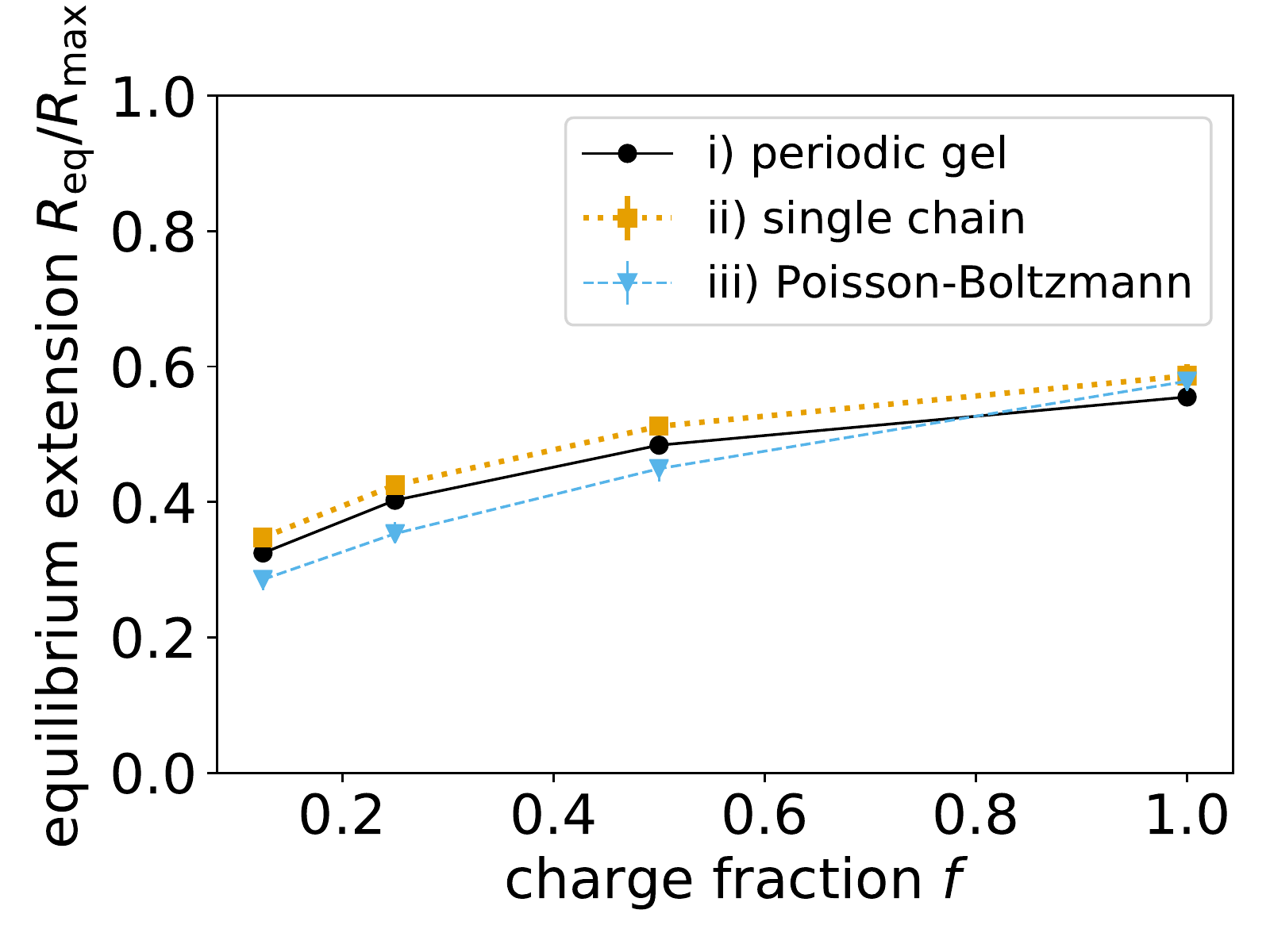}
\includegraphics[width=.8\columnwidth]{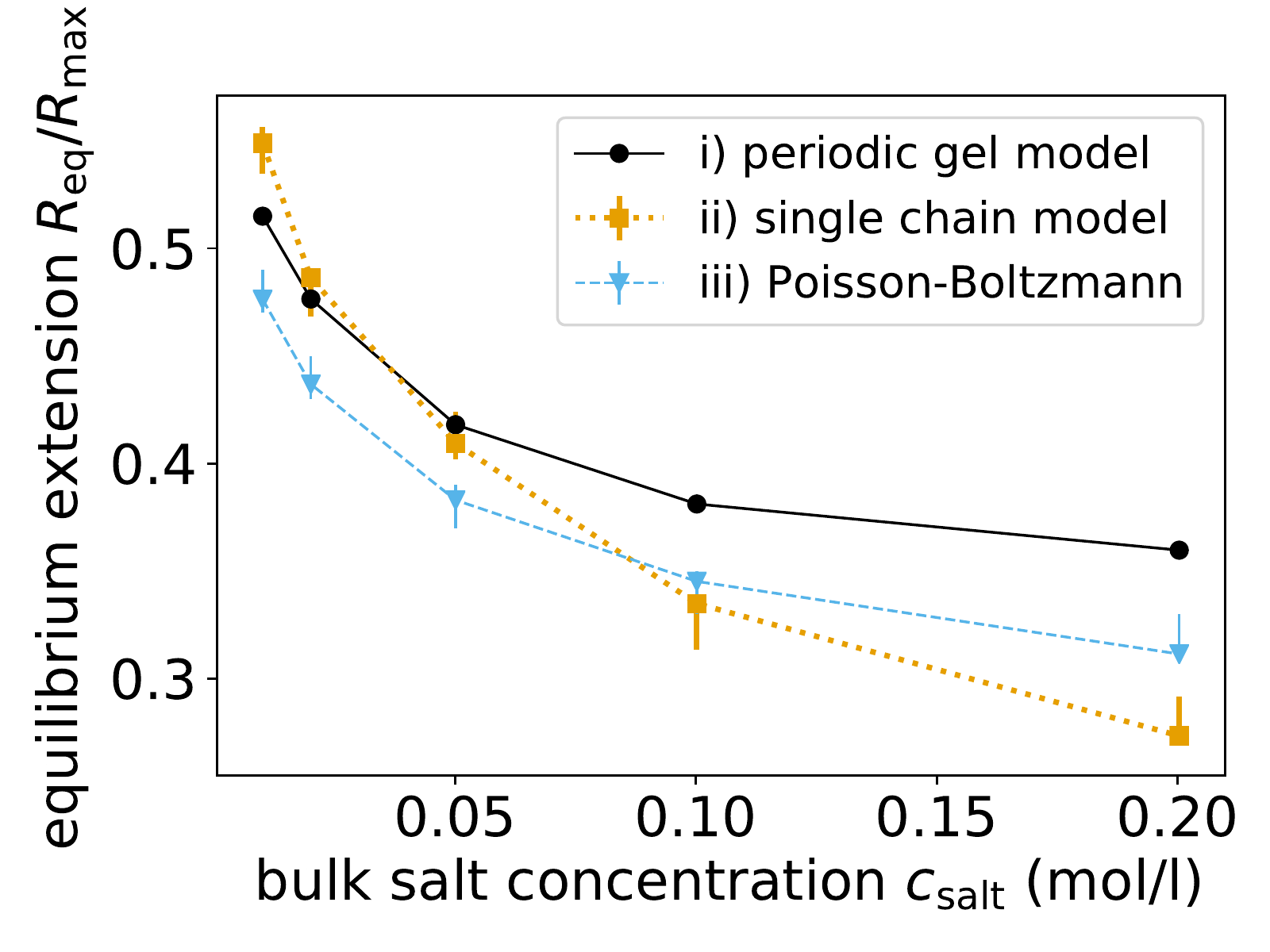}
\caption{\label{fig:params}(Color online) Comparison between the three
  models.  The equilibrium swelling length $\Req$ as a function of a)
  the charge fraction $f$ along the gel polymer backbone for
  $\cs=\SI{0.01}{\mole\per\liter}$ and $N \approx 80$;
  b) the salt bath
  concentration $\cs$ for $f=0.5$ and $N\approx 60$. 
  As can be seen in Fig. \ref{fig:correlation} the quality of the predictions of the single-chain model and the PB model for other parameter combinations are also very good.}
\end{figure}

The first model for complexity reduction is the mean-field single-chain
model. Like in the cylindrical cell model used to describe
solutions of polyelectrolytes~\cite{katchalsky53a,deshkovski01a,antypov06a}, we propose constructing the many-body partition function of a
periodic gel as a suitable product of individual cylindrical cells
containing a single polyelectrolyte chain with added salt. Since the
main physical principle is the balance between the polyelectrolyte
chain tension and the remaining pressure contributions (mainly the ionic ones), these
cylindrical cells have an axial length chosen such as to represent the
polymer chain extension between gel crosslinks. Like in the periodic
gel model, we perform MD simulations for a single-chain in cylindrical
confinement allowing explicit salt ion pairs to enter the cell volume
and reach chemical equilibrium with an external reservoir~\cite{frenkel02b}.  For a perfect affine
compression of a (fully stretched) tetrafunctional gel (built in a
diamond cubic lattice) the volume per chain is given by:
$V_\text{chain}=\Rend^3/A$, with the geometrical
prefactor $A=\sqrt{27}/4$~\cite{kosovan15a}.
Using the volume of the cylindrical cell as the volume per chain, we arrive at a
constant aspect ratio $\Rout/\Rend = 1/\sqrt{\pi A} \approx
0.49$, where $\Rout$ denotes the radius of the cylindrical cell, see
Fig.~\ref{fig:schematic}. 
This is in contrast to the deformation of a pure isolated
chain, which does not occupy a cylindrical volume of
constant aspect ratio upon a stretching deformation. 
The simulation volume is completely defined by the length of the cylinder,
or equivalently $\Rend$ \footnote{In the single chain MD simulations the cylinder height is $\Rend+b$ (with the average bond length $b\approx 0.966\sigma$) in order to be able to satisfy PBCs.}. 
Note that the single chain under confinement sees its images in the axial direction (due to PBCs) 
which in a simplified way mimics the electrostatic environment in a gel, where the end of a single chain sees the next chain. 
For cylindrical
geometries under affine compression, the pressure inside the volume
is given by~\cite{antypov06a,antypov06b}:
\begin{align}
\Pin = \frac{1}{3} \Pcap + \frac{2}{3} \Pside,
\label{eq:Pin}
\end{align}
where the total internal pressure $\Pin$ is split into the two
contributions from, $\Pcap$, the cylinder end caps and, $\Pside$, the side wall.
The latter is mainly dominated by collisions between mobile ions
and the boundary whereas the cap pressure is given as the
$(z, z)$ component of the pressure tensor $\Pcap=\Pi_{(z,z)}$:
\begin{align}
\Pi_{(z, z)}=\frac{\sum_{i} {m_{i}v_{i}^{(z)}v_{i}^{(z)}}}{V} + \frac{\sum_{j>i}{\vec{F}_{ij}^{(z)} \cdot \vec{r}_{ij}^{(z)}}}{V} +\Pi^\text{Coulomb}_{(z, z)}.
\end{align}
Here $V=\pi \Rout^2 \Rend$ is the \emph{effective available volume}, 
$m_{i}$ ($\vec{v}_i$) is the
mass (velocity) of particle $i$, and $\vec{F}_{ij}$ ($\vec{r}_{ij}$) the
force (connection vector) between particles $i$ and $j$.
The last term represents the Coulomb
contribution to the pressure tensor and is calculated according to Ref.~\cite{essmann95a}. 
The side contribution $\Pside$ is obtained directly by measuring the average normal force on the constraint and dividing by its area.
Having expressions for $\Pcap$, $\Pside$ and $\Pres$, we determine the equilibrium volume using $\Pin(\Veq)=\Pres$.
To check the accuracy of the single-chain cell model
we compare our equilibrium chain extensions to the ones obtained via
the periodic-cell model, cf.~Fig.~\ref{fig:params} of
Ref.~\cite{kosovan15a}. We will discuss the results
after describing the second mean-field approximation.

Since the single-chain cell model uses a cylindrical cell a further
reduction of the model complexity is to construct an
adapted PB description of the polyelectrolyte in the salt solution.
As before, the PB model uses a semi-infinite cylinder having an
external radius $\Rout$ and conceptual length $\Rend$, as shown in
Fig.~\ref{fig:schematic}. The electrostatics for an infinite rod is 
solved, again mimicking in a simplified way the electrostatic situation in a gel.
We also depict how the explicit
single chain is now modeled as a penetrable concentric charged rod of
radius $a$, characterized by a prescribed charge distribution.

The radius $a$ of the charged rod is chosen such that the polymer
density implies the same average distance from the end-to-end
vector as obtained from single-chain MD simulations. This amounts to finding
the length scale equivalent to the tensional blob size. The numerical
value is obtained via fitting a second degree polynomial
$a_\text{MD}(\Rend) = N \sigma \left( C_1 (\Rend / (N \sigma))^2 +
  C_2(\Rend /(N \sigma)) + C_3 \right)$, with $C_1=-0.17$, $C_2=0.14$,
$C_3=0.03$ based on our single-chain MD data.  Enforcing
$\langle r \rangle =\int_V p(\vec{r}) r\,\mathrm{d}V
\overset{!}{=}a_\text{MD}$ we obtain for the rectangular distribution
function $a=3/2 a_\text{MD}$.  The monomer charges thus have a
probability density $p(\vec{r})=\mathcal{N} H(-(r-a))$, where $H(x)$
is the Heaviside function and $\mathcal{N}$ a normalization such that
$p(\vec{r})$ is a probability density.  For a strong
polyelectrolyte, the charge is homogeneously distributed in the rod
with $\rho_f(\vec{r})= - N f e_0 p(\vec{r})$.

The Poisson equation describes the electrostatic interaction and ionic
distribution in the system:
\begin{align}
    \nabla^2\psi= -\frac{1}{\epsilon_0\epsilon_\mathrm{r}} \left( \sum_i q_i c_i(\vec{r}) +\rho_f(\vec{r}) \right).
\label{eq:PB}
\end{align}%
The number densities of the ions $c_i$ are related to the charge
densities via $\rho_i(\vec{r})=q_i c_i(\vec{r})$ given by
standard PB theory (with the bulk potential
$\psi^\mathrm{b}=0$):
\begin{equation}
c_i(r)=c_i^\mathrm{b} \exp \left(-\frac{q_i \psi(r)}{\kT} \right).
\end{equation}
Water is modeled implicitly via a relative dielectric
permittivity of $\epsilon_\mathrm{r} \approx 80$.

The PB pressure inside the cell $\Pin$ has two contributions:
1) The combined ideal and Maxwell pressure~\cite{trizac97a} which yields
\begin{align}
\Pside&= \kT  c(\Rout),\\
    \Pcap^\text{ions}&=\kT \langle c \rangle_z +\frac{\epsilon_0\epsilon_\mathrm{r}}{2} \langle E_r^2 \rangle_z,
\end{align}
where $E_r=-\partial_r \Psi(r)$ is the electric field in radial direction and $\langle \mathcal{A} \rangle_z=\int_0^{2\pi} \int_0^\text{$\Rout$} r \mathcal{A}(r) dr /(\pi \Rout^2)$ denotes the average over all radii. 
And 2) the stretching pressure $\Pcap^\mathrm{str}$ (acting only on the cap) which we define to be the pressure due to confinement $P^\mathrm{conf}$ (in the spirit of~\cite[p.\,115]{rubinstein03a}) minus the tensile stress $\sigma^\mathrm{chain}$ of the chain in order to
  have a finite extension for a neutral polymer gel:
  \begin{equation}
    \begin{split} &\Pcap^\mathrm{str}(\Rend)=P^\mathrm{conf}-\sigma^\mathrm{chain} \\
                         &=\frac{1}{\pi \Rout^2} \left(\frac{k_BT}{b} \frac{R_0^3}{\Rend^3} \mathcal{L}^{-1}\left(\frac{R_0}{R_\mathrm{max}}\right)  -\frac{k_B T}{b} \mathcal{L}^{-1}\left(\frac{\Rend}{R_\mathrm{max}}\right) \right),
    \end{split}
    \label{eq:Pstr}
  \end{equation}
  where $\mathcal{L}^{-1}$ is the inverse Langevin function.
  The stretching pressure is constructed such that
  $\Pcap^\mathrm{str}(R_0)=0$, where $R_0=1.2 b N^{0.588}$ is the
  average end-to-end distance of an unconfined neutral chain
 ~\cite{kosovan15a}. For a neutral gel, only the stretching pressure
  will determine the swelling equilibrium $\Pin(\Req)=\Pres$ which
  is found at $\Req=R_0$. The added confinement pressure dominates at low extensions.

Therefore the pressure inside the gel can be obtained via Eq. \eqref{eq:Pin} where the cap pressure is given by $\Pcap=\Pcap^\text{ions}+\Pcap^\mathrm{str}$. 
Applying the equilibrium condition $\Pin(\Req)=\Pres$, we obtain
the equilibrium end-to-end distance $\Req$. All equations were
solved with a finite element solver~\cite{comsol12a}.

\begin{figure}
\includegraphics[width=.8\columnwidth]{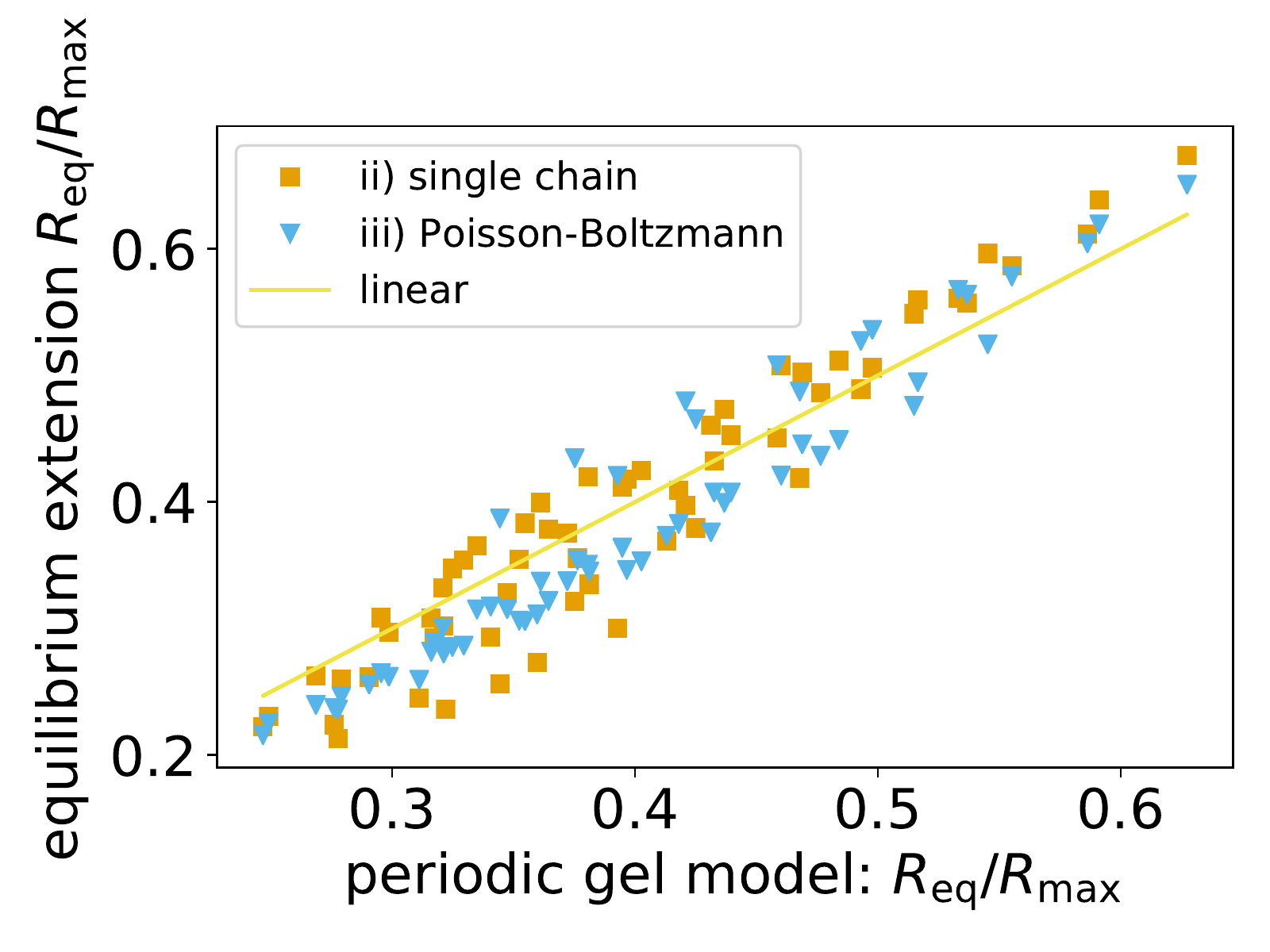}
\caption{\label{fig:correlation}(Color online) The swelling equilibria
  of the single-chain cell model and the PB model compared to the 
  periodic gel simulations.  The results are compared for a
  wide exploration of in total 60 parameter combinations with
  $N \approx 40, 60, 80$, $f \in \{0.125,0.25,0.5,1\}$ and
  \mbox{$\cs \in \{0.01,0.02,0.05, 0.1, 0.2\}$
    \si{\mole\per\liter}}. The linear function has the form
  $y(x)=x$. }
\end{figure}

In the following, we compare the obtained swelling equilibria for both
models to the periodic gel model for different charge
fractions, chain lengths, and reservoir salt concentrations. 
For selected parameters, Fig.~\ref{fig:params} demonstrates that for all models the gel swells:
i) more with increased charge fraction $f$;
%
%
ii) less with higher salt concentration in the reservoir $c_s^b$,
in good agreement with the data of the periodic gel model.
Both models also work for Manning parameters larger than unity, contrary
to the Katchalsky model~\cite{kosovan15a}. 
Further, at high charge fractions ($f>0.5$) they show better agreement with the periodic gel model than the
self-consistent field theory presented in
Ref.~\cite{rud17a}.
The PB model has basically the same accuracy as the single-chain MD
model for the selected parameter regions. Enlarging the comparison across a wide parameter range to
all available data, yielding 60 data
points, Fig.~\ref{fig:correlation} shows an excellent agreement of
both models
against the periodic gel model used here as the reference
standard.
Our PB model has the known limitations~\cite{andelman95a,deserno00a}:
multivalent ions, high charge densities
(e.g. at high compressions of the gel) or high ionic concentrations lead to deviations due to neglecting ionic
and excluded volume correlations which also exist in polyelectrolyte gels~\cite{yin09a} or charged
rod systems~\cite{deserno00a,deserno01b}. However, these limitations do not apply to our single chain MD model.

%

%
%
We now generalize the PB model to account for weak
groups similar to using the charge regulating boundary
condition~\cite{ninham71a}. However, we use charge regulation as a way to determine the space charge density
of the penetrable rod. For a weak polyelectrolyte where monomers
may be neutral or charged (\ce{HA <=> A- + H+}) the titratable
monomers (\ce{A-} or \ce{HA}) are distributed with $p(\vec{r})$
resulting in a concentration $c_0(\vec{r}) = N p(\vec{r})$. The dissociation constant is given by
$K_\mathrm{a}=c(\ce{A-})c(\ce{H+})/c(\ce{HA})=10^{-4}\si{\mole\per\liter}$.
Chemical equilibrium results in:
\begin{equation}
    c(\ce{A-},\vec{r})=\frac{c_0(\vec{r}) K_\mathrm{a}}{c^\mathrm{b}(\ce{H+})\exp(-e_0 \psi(\vec{r}) / (\kT ))+K_\mathrm{a}},
\end{equation}
and therefore $\rho_f(\vec{r})=-e_0 c(\ce{A-},\vec{r})$. In
the case of charge regulation we also explicitly model pH (while
neglecting any small \ce{OH-} concentration). Bulk charge neutrality
implies that the sum of the product of all species multiplied with
their valency needs to be zero (or equivalently):
\begin{equation}
    c^\mathrm{b}(\ce{H+})+c^\mathrm{b}(\ce{Na+})=c^\mathrm{b}(\ce{Cl-})
\end{equation}

Note that the bulk salt concentration
$c_\mathrm{salt}^\mathrm{b}$ is related to
$c^\mathrm{b}_{\ce{Na+}}=c_\mathrm{salt}^\mathrm{b}$ and
$c^\mathrm{b}_{\ce{Cl-}}=c_\mathrm{salt}^\mathrm{b}+c^\mathrm{b}_{\ce{H+}}$ to ensure charge neutrality.

\begin{figure}
\includegraphics[width=.8\columnwidth]{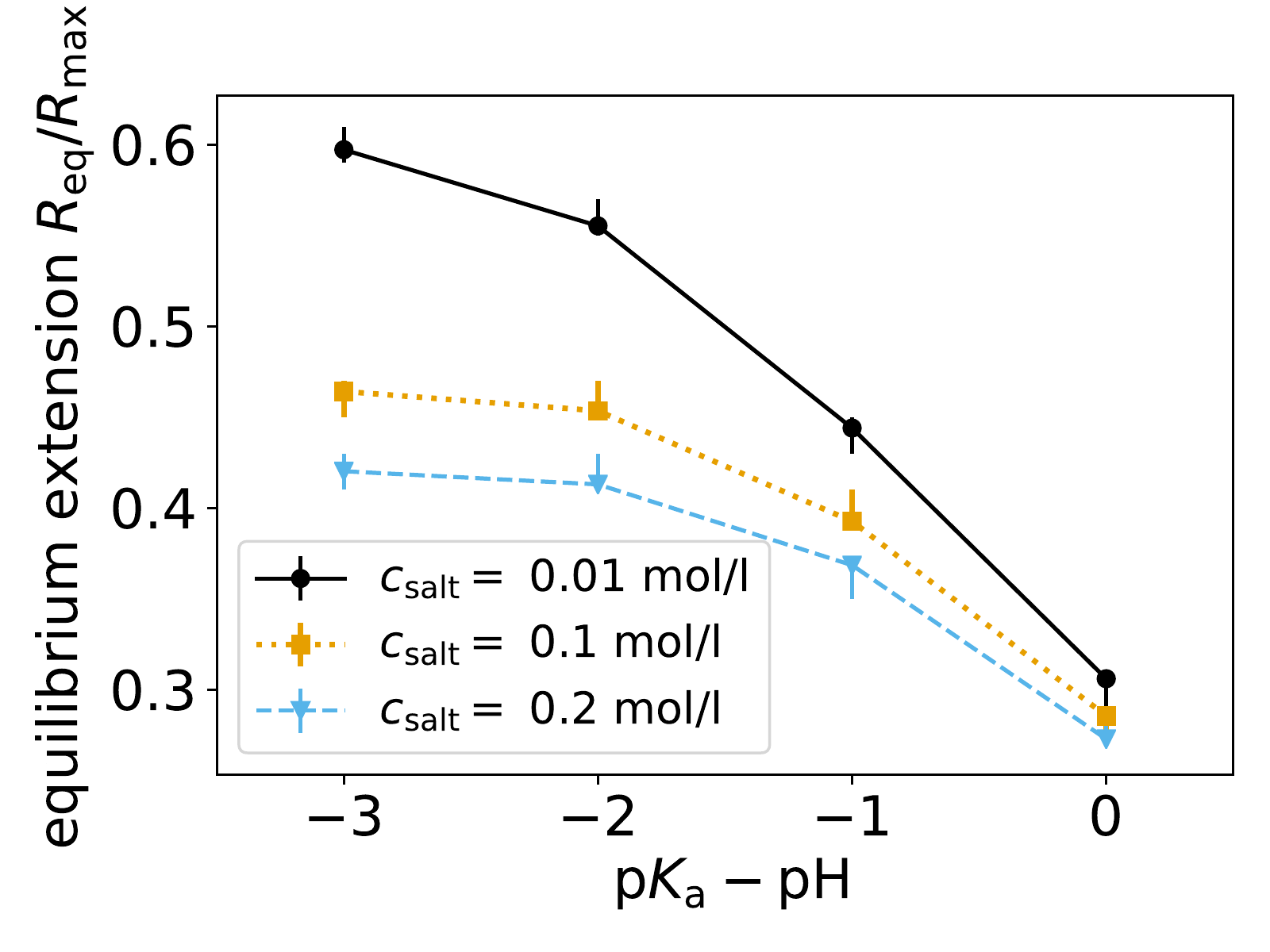}
\caption{\label{fig:weak}(Color online) The swelling equilibria as a
  function of $\pKa-\pH$ for different salt concentrations
  $\cs \in \{0.01, 0.1, 0.2 \}$ \si{\mole\per\liter} for $N =59$ and $K_\mathrm{a}=10^{-4}\si{\mole\per\liter}$.}
\end{figure}

In Fig.~\ref{fig:weak}, we show the equilibrium extension $\Req$ as a
function of the $\pKa-\pH$. The gel swells less with lower
$\pH=-\log_{10}(c^\mathrm{b}(\ce{H+})/\si{(\mole\per\liter)})$
($\pKa-\pH$ becoming larger) since the acid becomes less dissociated
(less charged). The gel also swells less with
higher salt concentration due to increased screening and a higher
pressure exerted by the salt reservoir. These findings are in good
qualitative agreement with experiments~\cite{katchalsky49b,
  tanaka81a}. 

In summary, we presented two successively coarsening mean-field models
that can predict swelling equilibria for charged macrogels.
The first one, the single-chain cell model, is a charged
bead-spring model with explicit salt ions confined within a
cylindrical cell which can undergo affine volume changes.  The
computational cost for solving the single-chain cell model is at least
an order of magnitude lower than for the periodic gel model.
In the next model we replaced the charged single-chain and all ions by
suitable charge distributions and use the PB framework to derive the
equilibrium cylindrical cell length. This model can be solved
numerically, i.e., with standard finite element solvers, and is yet at
least another order of magnitude faster than the single-chain cell
model. Since both models can predict the swelling equilibria
in similar good agreement for a wide parameter range with the more elaborate periodic
gel model, we can use the
extremely efficient PB model for predictions about gel swelling equilibria.
The PB model was further generalized to account for charged gels
containing weak groups. We find that the gel behavior under different
$\pKa-\pH$ conditions as well as salt reservoir concentrations
qualitatively agrees with experimental findings and theoretical
expectations.

\section{Acknowledgments}
  The authors acknowledge inspiring discussions with T.
  Richter and want to thank G. Rempfer and F. Weik for fruitful
  discussions regarding the pressure. Funding from the DFG through the
  SFB 716 and Grants HO 1108/26-1 and AR 593/7-1 is gratefully
  acknowledged.


\bibliographystyle{unsrt}
\bibliography{icp.bib}

\end{document}